# SIAMESE X-VECTOR RECONSTRUCTION FOR DOMAIN ADAPTED SPEAKER RECOGNITION

*Shai Rozenberg, Hagai Aronowitz, Ron Hoory*

IBM Research AI – Haifa, Israel

{shayr, hagaia, hoory}@il.ibm.com

## Abstract

With the rise of voice-activated applications, the need for speaker recognition is rapidly increasing. The x-vector, an embedding approach based on a deep neural network (DNN), is considered the state-of-the-art when proper end-to-end training is not feasible. However, the accuracy significantly decreases when recording conditions (noise, sample rate, etc.) are mismatched, either between the x-vector training data and the target data or between enrollment and test data. We introduce the Siamese x-vector Reconstruction (SVR) for domain adaptation. We reconstruct the embedding of a higher quality signal from a lower quality counterpart using a lean auxiliary Siamese DNN. We evaluate our method on several mismatch scenarios and demonstrate significant improvement over the baseline.

## 1. Introduction

Speaker recognition is a popular task, useful for processing various forms of communication, including telephone, broadcast, multimedia, and meeting data, as well as user authentication on mobile devices, smart home devices, and AI assistants.

The x-vector [1], an embedding approach for speaker verification based on a deep neural network (DNN), is considered the state-of-the-art embedding method for speaker verification. In addition, x-vector-based systems outperform other systems in a wide range of setups [2]. However, when given a sample from a different domain that was used to train the x-vector extractor, its accuracy significantly decreases. The different domain can be characterized by a different recording means, for example telephone versus studio, disturbances such as noise and additional mismatched conditions. In this paper, we explore the x-vector performance under such mismatched conditions. We explore three such mismatches: "noise", where random noise is added to the enrollment and/or test data; "sample rate", where data is sampled at either 16 KHz or 8 KHz; and "telephone", where we used a telephone simulator to simulate telephone speech. We chose these scenarios as we consider them most valuable to speaker recognition applications. For each scenario, we refer to the modified data as the lower quality data and to the original data as the higher quality data. We analyze the three mismatch scenarios to test speaker recognition performance where the enrollment data matches the x-vector high-quality training data, but the test data is of low quality. We also analyzed the "noise" and "telephone" scenarios to test a mismatch between high-quality x-vector training and low-quality target data (both enrollment and test data).

Our empirical study shows that the x-vector framework is very sensitive to the mismatches explored. To increase the x-vector robustness to lower quality data, we present the Siamese x-vector reconstruction (SVR) method in which we use an efficient Siamese DNN [3] trained to reconstruct the higher quality x-vectors on external data. The SVR method is novel in the context of embedding reconstruction and speaker recognition and has three main contributions. First, it dramatically improves the performance on low quality data without having to retrain the x-vector extractor on that data; this is a long and expensive process that requires the availability of the training data. Second, it allows overcoming quality differences between the enrollment and test data. Third, when using pre-trained x-vector models available in the KALDI [4] framework, two versions are available: an 8 KHz telephone version and a 16 KHz version, with the latter being more accurate and constantly improving. Our method allows using the more accurate 16 KHz x-vector model on 8 KHz and telephone data.

The rest of the paper is organized as follows: Section 2 includes a survey of related papers and compares methods; Section 3 describes the SVR method; Section 4 presents the performance for the three scenarios discussed; and Section 5 concludes and proposes future work.

## 2. RELATED WORK

In this paper, we follow the guidelines of the embedding framework for speaker recognition, in which speaker recognition is carried out by first embedding the speech utterance into a fixed-dimensional vector, and then scoring the similarity between the enrollment and test embeddings. Currently, the most popular speaker embedding methods are i-vectors [5] and x-vectors [1]. To determine whether two given embeddings originate from the same speaker, a probabilistic linear discriminant analysis (PLDA) model [6] is usually used.

The current state-of-the-art embedding for speaker recognition is the x-vector, which was introduced by Snyder et al. [1]. They used the intermediate activation layer of a DNN trained for speaker classification on a large amount of data, naming the embedding "x-vector". The x-vector currently has two open-sourced[1] versions that differ from one another based on the training data used. The 8 KHz telephone version was trained on more than 4K speakers [7] and the 16 KHz version was trained on more than 7K speakers from VexCeleb2 [8]. Since published, several improvements for the x-vector system have been suggested. Okabe et al. [9] replaced the average pooling done in the x-vector DNN with attention-based pooling [10]. Their method was able to improve the vanilla x-vector system, but was only tested for the 8 KHz system. Garcia et.al. [11] optimized the x-vector system by including an angular softmax [12] to score two vectors based on their cosine similarity. They demonstrated state-of-the-art results for a 16 KHz evaluation of speakers in the wild (SITW) [13].

---

[1] Models can be found in http://kaldi-asr.org/models.html

While the x-vector achieves state-of-the-art accuracy on datasets such as SITW, its accuracy degrades significantly in cases of mismatched conditions. The idea of embedding-level modification is tackled by Kheder et,al. [14]. They modified the i-vector embedding of a noisy utterance by estimating the clean i-vector from the noisy version, based on assumptions about the noise distribution. While this approach exhibits a significant improvement compared to the vanilla i-vector, this method is limited to noisy data. Other researchers attempted to recover a higher quality version of the audio, such as Plchot et al. [15] who used a neural network to denoise the audio, and Kuleshov et al. [16] who used an encoder-decoder-based DNN to perform audio super resolution.

In contrast to the methods reviewed thus far, other methods attempt to overcome domain mismatch given existing embeddings. Score Normalization (SN) takes a single trial (enrollment, test recordings, and corresponding matching score) and normalizes the score according to the score distributions of the enrollment/test recordings with respect to an imposter dataset [17]. PLDA Adaptation (PA) [18] uses an unlabeled adaptation dataset to modify the PLDA model used to score the samples. This method exhibits significant improvements when dealing with channel and language mismatches. The PA method uses both original and adapted PLDA models with a parameter $\alpha$ to weigh them.

Our method can be used in conjunction with other methods. Specifically, we combine SN with our method for further improvement, as reported in Section 4.

## 3. SIAMESE X-VECTOR RECONSTRUCTION

Our goal is to reconstruct a higher quality x-vector from a low-quality counterpart, without having to retrain the x-vector embedding network. Such retraining is not only an expensive procedure, but may also be infeasible due to lack of data or skills. Furthermore, we want to be able to score a high-quality enrollment utterance against a low-quality test utterance. Therefore, we propose modifying the x-vector directly, rather than the signal. To that end, we begin by creating a DNN that receives a low-quality x-vector and is trained to reconstruct the high-quality x-vector. Formally, let $F_\theta$ denote a neural network with parameters $\theta$ whose inputs and outputs have the same dimension as the x-vector embedding, $x \in R^d$, $F_\theta(x) \in R^d$. Following that, we utilize the knowledge of the task at hand and aim to have x-vectors of the same speakers from two different utterances be more similar than those of two different speakers. Hence, we use the reconstruction DNN as the basis of a Siamese DNN [3]. To do that, we create two identical, weight-shared, instances of $F_\theta$. We then use the cosine similarity between the reconstructed x-vectors of two given inputs, setting the target value to 1 if the two input low-quality embeddings belong to the same speaker, and 0 otherwise. We call our method the Siamese X-Vector Reconstruction (SVR).

Formally, let $x_1^l, x_2^l$ denote two low-quality x-vectors from speakers $c_1$, $c_2$ respectively and $x_1^h, x_2^h$ their high-quality counterparts. We train $F_\theta$ using a summation of the two reconstruction losses and the cosine similarity loss denoted by L:

$$L = ||F_\theta(x_1^l) - x_1^h||^2 + ||F_\theta(x_2^l) - x_2^h||^2 + \frac{F_\theta(x_1^l) \cdot F_\theta(x_2^l)}{||F_\theta(x_1^l)||||F_\theta(x_2^l)||} - \delta_{c_1,c_2} \quad (1)$$

Where,

$$\delta_{c_1,c_2} = \begin{cases} 1 & \text{if } c_1 = c_2 \\ 0 & \text{otherwise} \end{cases}$$

*Figure 1: SVR system with two input, low-quality, x-vectors fed to the DNN and a reconstruction loss coupled by a cosine similarity loss used to train $F_\theta(x)$*

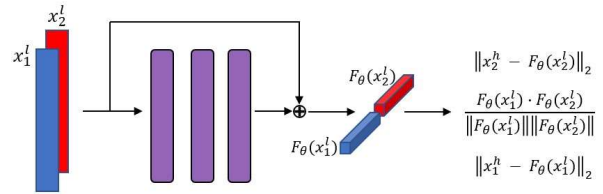

To reconstruct an x-vector during inference, one simply uses the output of $F_\theta$ without using the Siamese training scheme. The architecture we chose for $F_\theta$ is a simple three-layer, fully connected DNN. As a result, our method is highly efficient and can be utilized for several types of mismatch conditions and other forms of embedding (not limited to x-vectors). We trained our model on the test portion of VoxCeleb2 [8], which was not used during the training of the x-vector model. This form of training took just a few hours using a standard single GPU machine, as opposed to a few days that it takes to train the x-vector model. Our method doesn't require any additional data from the test or enrollment domain. However, as such data can be provided in several applications, we evaluated the possibility of combining our method with score normalization.

## 4. EXPERIMENTS

For the evaluation, we consider two forms of mismatch. The first is where the enrollment data mismatches the test data, with the enrollment data being of higher quality. For instance, the enrollment data is sampled at 16 KHz, while the test is sampled at 8 KHz. That's a feasible scenario, as the enrollment data is collected once and as such can be obtained rigorously. We refer to this form of mismatch as "Original Enrollment", because only the test data is of reduced quality. The second form is where the enrollment and test data share the same level of quality, which is different from that used for training the x-vector network. For instance, both enrollment and test data are sampled at 8 KHz, but we wish to use the more accurate, 16

KHz x-vector model, which was trained on more speakers and is constantly evolving. We evaluated both forms under three mismatch scenarios: noise, sampling rate, and telephone simulation. For each scenario, we trained an SVR model to overcome the mismatch. We evaluated our performance on the common benchmark SITW [13], a 16 KHz dataset consisting of several utterances per speaker with a few hundred speakers, using the core-core partition for evaluation. In addition to comparing SVR to the vanilla x-vector, using the development set, we compare SVR to the score normalization (SN) and the PLDA adaptation (PA) methods and to a combination of the SVR and the SN methods.

We follow the guidelines defined by the NIST 2019 Speaker Recognition Evaluation: CTS Challenge and report our results as the equal error rate (EER) and minimum of the normalized detection cost function (minDCF): $DCF(\theta;\beta) = P_{FN}(\theta) + \beta \times P_{FP}(\theta)$, where $P_{FN}$ and $P_{FP}$ are the false negative and false positive ratios per threshold $\theta$, respectively, and $\beta$ is a hyper-parameter. The DCF is averaged over two values of $\beta$: *99* and *199*.

### 4.1. Baselines

We began by evaluating the performance of the baseline open-sourced models on the original 16 KHz SITW and a down-sampled 8 KHz version. Table 1 shows the EER and minDCF using the two x-vector extractors. The first row corresponds to the 16 KHz x-vector extractor used on the SITW dataset and the second row to the 8 KHz extractor used on a down-sampled 8 KHz version of SITW. The 16 KHz model significantly outperforms the 8 KHz model. We therefore consider the 3.5% EER and 0.38 minDCF as lower bounds for the remaining experiments.

*Table 1: Baseline results, using the 8 KHz and 16 KHz x-vector models on the SITW dataset*

| Enroll Sample Rate | Test Sample Rate | EER (in %) | minDCF |
|---|---|---|---|
| 16 KHz | 16 KHz | 3.5 | 0.38 |
| 8 KHz | 8 KHz | 7 | 0.59 |

### 4.2. Noise Mismatch

The first scenario we examined is the noise mismatch. In this scenario, we created a noisy version of the SITW 16 KHz dataset. We added a random combination of white, pink, and brown noise resulting in a dataset with an estimated signal-to-noise ratio (SNR) of between 8 and 20 dB. An SVR model was trained on a noisy version of the VoxCeleb2 test set. The results for this experiment are presented in Table 2

*Table 2: Noise mismatch results on the SITW dataset. The x-vector extractor was trained on clean speech. Test sessions are always noisy.*

| | Original Enrollment | | Noisy Enrollment | |
|---|---|---|---|---|
| | EER (in %) | minDCF | EER (in %) | minDCF |
| Baseline | 5.4 | 0.52 | 7.3 | 0.82 |
| SN | 5.3 | 0.49 | 7.6 | 0.58 |
| PA | 5.7 | 0.53 | 7.5 | 0.64 |
| SVR | 3.9 | 0.42 | 4.1 | **0.45** |
| SVR+SN | **3.6** | **0.4** | **3.9** | **0.45** |

We first note from the baseline row that the added noise had a profound impact on the performance. Using a simple, defined noise structure, it increased the EER by 55% and 108% for the two scenarios, respectively. Second, we observe that our method, SVR, significantly outperforms the other methods. The best results were achieved by combining SVR with score normalization, which obtained EERs of 3.6 and 3.9, close to the lower bound of 3.5 achieved on the clean version of the data.

The PLDA adaptation technique used throughout the experiments is dependent on the choice of the weighting parameter. The EER and corresponding minDCF presented were the lowest achieved with various values tried, as illustrated in Figure 2. Even with this "cheating" experiment, the PLDA adaptation did not improve the results. It also did not improve results for the other scenarios.

*Figure 2: PLDA adaptation for various values of α for the noise mismatch scenario with noisy enrollment. The lowest EER achieved is 7.5%*

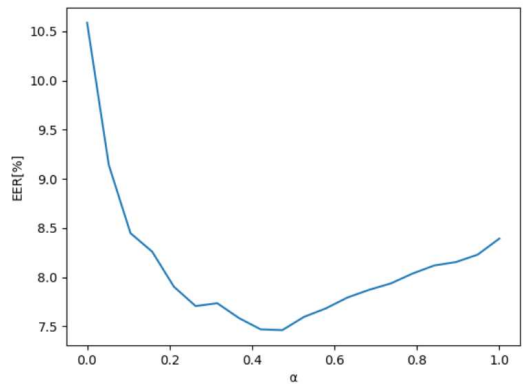

### 4.3. Sample Rate Mismatch

For the second scenario, we examine data obtained in a low sample rate. We down-sampled the SITW dataset and created a low-quality, 8 KHz version. As the 16 KHz x-vector extractor is more accurate, as evident from *Table 1*, we attempt to utilize it with our data by upsampling back to 16 KHz. For SVR, we train a DNN to reconstruct a 16 KHz x-vector from its 8 KHz counterpart. We evaluate the speaker verification performance given that the enrollment is of a higher quality (16 KHz) and given both the enrollment and test data were upsampled from 8 KHz.

*Table 3: Sample rate mismatch results on the SITW dataset. The test data is upsampled from 8 KHz to 16 KHz, while the enrollment is either 16 KHz (original) or upsampled from 8 KHz as well.*

| | Original Enrollment | | Upsampled Enrollment | |
|---|---|---|---|---|
| | EER (in %) | minDCF | EER (in %) | minDCF |
| Baseline | 7.6 | 0.75 | 8.8 | 0.75 |
| SN | 5.9 | 0.57 | 6.6 | 0.59 |
| PA  | 6.3 | 0.72 | 8.1 | 0.6  |
| SVR | 5.3 | 0.57 | 6   | 0.6  |
| **SVR+SN** | **4.7** | **0.55** | **5.7** | **0.58** |

From the results in Table 3 and Table 1, we note that, surprisingly, even when the enrollment data is given in higher quality than the test data (see Table 3, first two columns), it would be better to down-sample it to 8 KHz as the EER improves from 7.6 to 7 and the minDCF from 0.75 to 0.59 (Table 1 second row). Moreover, even though the 16 KHz x-vector model is considered superior, upsampling the data and using it yields much worse results (see Table 3, columns 3 and 4). In contrast, the SVR method shows improvement compared to the 8 KHz baseline, reducing the EER by 33%.

These results imply that when using SVR, one may use the higher quality x-vector model, which is more accurate and constantly improving, on lower quality data. This is preferable to matching the quality of the data between the enrollment and test to that of the lower quality, the second best option, which performs significantly poorer.

### 4.4. Phone Simulator

The third scenario we considered is one where the test and/or enrollment data are taken from a telephone channel. To create pairs of telephone and "clean" instances, we used a telephone simulator [19] and created a 8 KHz telephone version of VoxCeleb2 for the training of SVR and SITW for evaluation. We note that the telephone simulation mostly had a similar effect on the recognition performance as in the previous scenarios. Table 4 indicates that the SVR method significantly outperform the baselines, reducing the EER by 35% when using the original enrollment data and 16% when the enrollment is also acquired via a telephone channel.

*Table 4: Telephone simulation results on the SITW dataset. The test data is simulated as acquired via a telephone channel while the enrollment is either clean or simulated as well.*

|          | Original Enrollment | | Telephone Simulated Enrollment | |
| -------- | --------- | ------ | --------- | ------ |
|          | EER (in %) | minDCF | EER (in %) | minDCF |
| Baseline | 8         | 0.78   | 7.3       | 0.61   |
| SVR      | 5.6       | 0.61   | 6.3       | 0.63   |
| **SVR+SN** | **5.2** | **0.57** | **6.1** | **0.6** |

## 5. CONCLUSIONS AND FUTURE WORK

In this paper, we presented the Siamese x-vector reconstruction (SVR) method for improving speaker recognition under mismatch conditions. Our contribution is three-fold. We first significantly improve the performance on low-quality data compared to the vanilla x-vector model. Second, when dealing with quality differences between the enrollment and test data, our method exhibits better performance than downgrading the higher quality data. Third, and perhaps most important, our method enables using the higher quality 16 KHz x-vector model on 8 KHz data.

The three scenarios we presented (noise mismatch, sample-rate mismatch, and telephone simulation) had a significant impact on the performance of the vanilla x-vector extractor, revealing a serious shortcoming of this extractor in the case of domain mismatch.

The SVR can be generalized to any given embedding and cope with additional scenarios. As such, we intend to explore the use of SVR on such scenarios and embeddings, potentially extending it beyond the scope of the speaker recognition task, as the SVR addresses issues that are prominent in many research fields.